\documentclass[%
 reprint,
 superscriptaddress,
 amsmath,amssymb,
 aps,
 pra,
]{revtex4-1}

\usepackage{graphicx}
\usepackage{dcolumn}
\usepackage{bm}
\usepackage{hyperref}

\begin{document}

\preprint{APS/123-QED}

\title{Exploring dynamical quantum phase transition from pure states to mixed states through extended Su-Schrieffer-Heeger models}

\author{Kaiyuan Cao}
 \email{kycao@yzu.edu.cn}
\affiliation{College of Physics Science and Technology, Yangzhou University, Yangzhou 225002, China}

\author{Tianren Zhang}
\affiliation{College of Physics Science and Technology, Yangzhou University, Yangzhou 225002, China}

\author{Xiang-Ping Jiang}
 \email{2015iopjxp@gmail.com}
\affiliation{School of Physics, Hangzhou Normal University, Hangzhou, Zhejiang 311121, China}


\author{Jian Wang}
 \email{wangjian@yzu.edu.cn}
\affiliation{College of Physics Science and Technology, Yangzhou University, Yangzhou 225002, China}

\date{\today}

\begin{abstract}
  We investigate dynamical quantum phase transitions (DQPTs) in both pure and mixed states within the extended SSH model framework, focusing on the SSH-3 and SSH-4 variants, which differ in symmetry properties. The SSH-3 model, characterized by a chiral-like point symmetry rather than true chiral symmetry, supports robust localized edge states tied to its topological nature. Our results show that for pure states, DQPTs occur after quenches crossing the topological transition, even when the energy band gap remains open. For mixed states, DQPT behavior aligns with pure states at low temperatures but undergoes significant changes at higher temperatures, including the emergence of multiple critical times. In contrast, the SSH-4 model, which possesses chiral symmetry, features four distinct energy spectrum configurations. We find that pure-state DQPTs arise only when the quench starts from a gapless initial state and crosses the critical topological point. At finite temperature, mixed-state DQPTs persist at low temperatures only if the corresponding pure-state quench induces DQPTs, but they disappear at elevated temperatures. These findings elucidate the interplay between symmetry, topology, and temperature in governing DQPTs within generalized SSH models.
\end{abstract}

\maketitle


\section{Introduction}

Recent experimental investigations on ultra-cold atoms confined in optical lattices \cite{Bloch2008rmp, Lewenstein2012} have significantly advanced the exploration of non-equilibrium dynamics in isolated quantum systems \cite{Polkovnikov2011rmp}. A primary focus within this field is the temporal evolution of a quantum system following a sudden global quench, which can be readily performed in experiments and studied theoretically \cite{Aditi2018arcmp}. In these scenarios, the Loschmidt echo, indicating the overlap between the pre- and post-quench Hamiltonian eigenstates, plays a pivotal role \cite{Zurek2005prl}. The formal parallelism between the Loschmidt amplitude and the canonical partition function of an equilibrium system has led to the concept of dynamical quantum phase transitions (DQPTs), aiding in the comprehension of phase and phase transition concepts in non-equilibrium settings \cite{Heyl2013prl, Zvyagin2016ltp, Heyl2018rpp}.

The concept of dynamical quantum phase transitions (DQPTs) specifically elucidates the critical behaviors of the Loschmidt echo, $\mathcal{L}(t) = |\mathcal{G}(t)|^2$, in the non-equilibrium evolution of quantum systems. The Loschmidt amplitude $\mathcal{G}(t)$, quantifying the overlap between evolving and initial states, is expressed as $\mathcal{G}(t) = \langle\psi_{0}|e^{-iHt}|\psi_{0}\rangle$. To manifest DQPTs, the dynamical free energy density is defined as the rate function of the Loschmidt echo in the thermodynamic limit, $\lambda(t)=-\lim_{N\rightarrow+\infty}\frac{1}{N}\ln{[\mathcal{L}(t)]}$, which exhibits cusp-like singularities at critical times. Recent advancements have unveiled that DQPTs can shed light on the topological aspects of real-time quench dynamics \cite{Budich2016prb}. The DQPT has been extensively studied in various quantum systems, including integrable models (e.g. XY chains \cite{Vajna2014prb, Cao2022cpb, Porta2020scirep}, Kitaev honeycomb model \cite{Schmitt2015prb}), non-integrable models \cite{Karrasch2013prb, Andraschko2014prb, Heyl2014prl, Kriel2014prb, Sharma2015prb}, systems with long-range interactions \cite{Halimeh2017prb, Homrighausen2017prb, Obuchi2017prb, Valentin2017pre, Dutta2017prb, Bojan2018prl, Halimeh2020prr}, non-Hermitian systems \cite{Zhou2018pra, Mondal2022prb, Zhai2022prb, Mondal2023prb}, inhomogeneous systems \cite{Yang2017prb, Yin2018pra, Srivastav2019prb, Cao2020prb, Mishra2020JPhysA, Modak2021prb, Hoyos2022prb, Cao2023prb, Kuliashov2023prb}, and among others \cite{Karrasch2017prb, Lahiri2019prb, Mehdi2019prb, Huang2019prl, Zache2019prl, Jafari2019prb}. Besides the quantum quench, other nonequilibrium protocols, for instance, the linear ramp \cite{Divakaran2016pre, Sharma2016prb, Tatjana2016SciPost, Zamani2024JPhysCondens, Cao2024pra}, periodic Floquet systems \cite{Yang2019prb, Zamani2020prb, Shirai2020pra, Zhou2021jpcm, Jafari2021pra, Hamazaki2021nc, Zamani2022prb, Jafari2022prb}, and systems 
subject to noise \cite{Jafari2024prb}, have also attracted wide attention. Moreover, experimental validations of DQPTs have been achieved through trapped ions simulations \cite{Vogel2017naturep, Jurcevic2017prl, Chen2020pra, Muniz2020nature}, nuclear magnetic resonance quantum simulators  \cite{Nie2020prl}, quantum walks of photons \cite{Wang2019prl, Xu20209lightsa}, and spinor condensate simulations \cite{Tian2020prl}. Additionally, a similar definition of the dynamical phase transition, studying the asymptotic late-time steady state of the order parameters \cite{Barmettler2009prl, Eckstein2009prl, Sen2016prb}, also arouses a lot of interest. Two types of DQPTs have been found related to the long-range quantum Ising chain \cite{Bojan2018prl}.

Recent studies have intriguingly extended the theory of DQPTs to nonzero temperatures \cite{Abeling2016prb, Bhattacharya2017prbL, Heyl2017prbL}. By generating the Loschmidt amplitude for initial mixed state $\rho_{0}$ via the interferometric interpretation in topological systems after a quantum quench \cite{Bhattacharya2017prbL, Heyl2017prbL}, it is found that the DQPT is not eliminated but rather persists at relatively high initial temperatures. This finding highlights that DQPTs possess fundamentally different characteristics compared to quantum phase transitions, which are known to disappear at finite temperatures. It is important to note that the extension of the Loschmidt amplitude to nonzero temperatures is not unique. Lang and colleagues introduced a generalized Loschmidt amplitude for mixed states \cite{Lang2018prb} under the parallel transport condition \cite{Sjoqvist2000prl, Tong2004prl}, and similarly confirmed that DQPT is not destroyed at finite temperatures, even under sufficiently weak dissipation. The only difference between both generations of the Loschmidt amplitude is the phase factor when the state is pure and the dynamics are unitary. Notably, the Fisher zero points and rate function remain unchanged \cite{Lang2018prb, Tang2410.17940}. However, in our subsequent study of mixed-state DQPTs in the multi-band Kitaev model, we found that the presence of energy gaps between the system’s energy levels leads to the vanishment of DQPT at nonzero temperatures \cite{Cao2024jpcm}. This observation motivated us to further explore how different band structures influence mixed-state DQPTs.

In this paper, we focus on extended Su-Schrieffer-Heeger (SSH$-m$) models featuring multiple energy bands ($m > 2$). The general SSH$-m$ models show different bulk symmetries for odd and even $m$, so that we specifically examine two representative cases: the SSH-3 and SSH-4 models. The SSH-3 model exhibits a chiral-like point symmetry rather than a true chiral symmetry \cite{Anastasiadis2022prb}, whereas the SSH-4 model possesses genuine chiral symmetry \cite{Lee2022cjp}. We investigate the properties of DQPTs for both pure and mixed initial states by computing the rate function and the Pancharatnam geometric phases. Our results reveal that these different symmetries strongly affect the behavior of DQPTs. 

The paper is organized as follows: In Sec.~\ref{sec:model}, we introduce the generalized SSH model and derive the time-dependent state of the system after a quantum quench. In Secs.~\ref{sec:pure.ssh3}, we investigate the behaviors of DQPTs for both pure and mixed states in the SSH-3 model. In Sec.~\ref{sec:pure.ssh4}, we study the properties of the DQPTs for both pure and mixed states in the SSH-4 model. We conclude in Sec.~\ref{sec:conclusion}

\section{model}
\label{sec:model}

We consider a generalized version of the multi-band SSH model, referred to as the SSH$-m$ model, whose Hamiltonian is expressed as follows: 
\begin{equation}\label{eq:SSH-m}
    \begin{split}
        H & = -\sum_{n=1}^{N} (t_{1}|n,A\rangle\langle n,B| + \cdots + t_{m-1}|n,M-1\rangle\langle n, M|) \\
          & \quad -\sum_{n=1}^{N} t_{m}|n,M\rangle\langle n+1,A| + H.c.,
    \end{split}
\end{equation}
where $t_{1}, \cdots, t_{m-1}$ represent the intracell couplings, and $t_{m}$ denotes the intercell coupling. In contrast to the SSH model, the generalized SSH-m model reveals a richer variety of topological phase transitions \cite{Anastasiadis2022prb, Lee2022cjp, Martinez2019pra}. By imposing the periodic boundary condition, we obtain the bulk Hamiltonian as 
\begin{equation}\label{eq:Bloch.m}
    H_{k} = \left(
                \begin{array}{ccccc}
                    0 & t_{1} & 0 & \cdots & t_{m}e^{-ik} \\
                    t_{1} & \ddots & \ddots & \ddots & \vdots \\
                    0 & \ddots & 0 & \ddots & 0 \\
                    \vdots & \ddots & \ddots & \ddots & t_{m-1} \\
                    t_{m}e^{ik} & \cdots & 0 & t_{m-1} & 0 \\
                \end{array}
            \right),
\end{equation}
where $k$ is the lattice momentum.

We now consider the quantum quench by suddenly changing the intercell coupling parameter $t_{m}$. The system is initialized in a pure state $|u_{k\mu}^{i}\rangle$, which satisfies
\begin{equation}
    H_{k}(t_{m}^{i})|u_{k\mu}^{f}\rangle = \varepsilon_{k\mu}^{i}|u_{k\mu}^{i}\rangle, \mu = 1, \cdots, m.
\end{equation}
At $t = 0$, the intercell coupling parameter is changed from $t_{m}^{i}$ to $t_{m}^{f}$. The time-evolved state following the quench is then given by
\begin{equation}\label{eq:time-evolved.state}
    |\psi_{k}(t)\rangle = e^{-iH_{k}(t_{1}^{f})t}|u_{k\mu}^{i}\rangle.
\end{equation}
To calculate Eq.~(\ref{eq:time-evolved.state}), we can write the initial state as a linear superposition of eigenstates of the post-quench Hamiltonian, which yields
\begin{equation}\label{eq:superposition}
    |u_{k\mu}^{i}\rangle = \sum_{\nu = 1}^{m}p_{k\nu}|u_{k\nu}^{f}\rangle, \quad p_{k\nu} = \langle u_{k\nu}^{f}|u_{k\mu}^{i}\rangle.
\end{equation}
Substituting Eq.~(\ref{eq:superposition}) into (\ref{eq:time-evolved.state}), we obtain the time-evolved state as 
\begin{equation}
    |\psi_{k}(t)\rangle = \sum_{\nu = 1}^{m}p_{k\nu}e^{-i\varepsilon_{k\nu}^{f}t}|u_{k\nu}^{f}\rangle.
\end{equation}
We thus obtain the Loschmidt amplitude $\mathcal{G}(t) = \prod_{k}\mathcal{G}_{k}(t)$ immediately with
\begin{equation}\label{eq:LA.k}
    \mathcal{G}_{k}(t) = \langle u_{k\mu}^{i}|\psi_{k}(t)\rangle = \sum_{\nu=1}^{m}|p_{k\nu}|^{2}e^{-i\varepsilon_{k\nu}^{f}t}.
\end{equation}
It is convenient to express the Loschmidt amplitude $\mathcal{G}_{k}(t)$ as
\begin{equation}
    \mathcal{G}_{k}(t) = r_{k}(t)e^{i\phi_{k}(t)}
\end{equation}
in the polar coordinate, so that the Loschmidt echo can thus be obtained by
\begin{equation}
    \mathcal{L}(t) = \prod_{k}\mathcal{L}_{k}(t), \quad \mathcal{L}_{k}(t) = r_{k}^{2}(t).
\end{equation}
In the thermodynamic limit, the DQPTs are identified by the cusp-like singularities of the dynamical free energy density, which is defined as the rate function of the Loschmidt echo:
\begin{equation}
    \lambda(t) = -\lim_{N\rightarrow\infty}\frac{1}{N}\ln{|G(t)|^{2}} = -\int_{-\pi}^{\pi}\frac{dk}{2\pi}\ln{r_{k}^{2}(t)}.
\end{equation}
Since $r_{k}(t)$ is well-defined, the singularities of the rate function $\lambda(t)$ occur only for $r_{k}(t) = 0$, which is known as the Fisher zeros of the Loschmidt amplitude. At the Fisher zeros of $\mathcal{G}_{k}(t)$, the associated phase $\phi_{k}(t)$ is ill-defined, which induces the dynamical topological phase transition \cite{Budich2016prb}. According to Berry's theory, the total phase $\phi_{k}(t)$ consists of a geometric phase $\phi_{k}^{G}(t)$, known as the Pancharatnam geometric phase, and a dynamical phase $\phi_{k}^{dyn}(t)$, in which the latter is contributed from the accumulation of the energy. In our model, the dynamical phase for the pure state can be obtained by
\begin{equation}
    \phi_{k}^{dyn}(t) = -\sum_{\nu=1}^{m}|p_{k\nu}|^{2}\varepsilon_{k\nu}^{f}t.
\end{equation}

In contrast, when considering the dynamics at finite temperature, the initial state is prepared as the thermal state described by a full-ranked density matrix $\rho_{0}$. The density matrix at time $t$ after the quench is given by $\rho(t) = e^{-iH(t)}\rho_{0}e^{iHt}$. Hence, a generalization of the Loschmidt amplitude to general density matrices we consider is defined as \cite{Heyl2017prbL, Bhattacharya2017prbL}
\begin{equation}\label{eq:LA.generalization}
    \mathcal{G}(t) = \mathrm{Tr}[\rho_{0}U(t)] = \prod_{k}\mathrm{Tr}[\rho_{0k}U_{k}(t)],
\end{equation}
where $U(t)$ is the time-evolution operator.
It is important to emphasize that the generalized Loschmidt amplitude proposed here is not limited by the parallel transport condition \cite{Sjoqvist2000prl, Tong2004prl, Lang2018prb, Tang2410.17940}.
In our model, the initial density matrices for every $k$ take the form
\begin{equation}\label{eq:density.matrix.init}
    \rho_{0k} = \sum_{\mu=1}^{m} f_{k\mu}|u_{k\mu}^{i}\rangle\langle u_{k\mu}^{i}|,
\end{equation}
where the probabilities $f_{k\mu} \in (0, 1]$ of the electron occupying $|u_{k\mu}^{i}\rangle$ parameterize the thermal state
\begin{equation}
    f_{k\mu} = \frac{e^{-\beta\varepsilon_{k\mu}^{i}}}{\sum_{\mu=1}^{m}e^{-\beta\varepsilon_{k\mu}^{i}}}.
\end{equation}
Here, $\beta = 1/T$ is the inverse temperature. For simplicity, we set $c = \hbar = k_{b} = 1$ in our paper. Substituting Eq.~(\ref{eq:density.matrix.init}) into Eq.~(\ref{eq:LA.generalization}), we immediately obtain the generalized Loschmidt overlap amplitude by
\begin{equation}\label{eq:gLA.k}
    \mathcal{G}(t) = \sum_{\mu=1}^{m}f_{k\mu}\sum_{\nu=1}^{m}|p_{k\nu}|^{2}e^{-i\varepsilon_{k\nu}^{f}t}.
\end{equation}
For a pure initial state $|u_{k\mu}^{i}\rangle$, i.e. $f_{k\mu} = 1$, Eq.~(\ref{eq:gLA.k}) returns back to Eq.~(\ref{eq:LA.k}). 

Similarly, we can also express the generalized Loschmidt amplitude in the polar coordinate, i.e. $\mathcal{G}_{k}(t) = R_{k}(t)e^{i[\Phi_{k}^{dyn}(t)+\Phi_{k}^{G}(t)]}$, where the dynamical phase for the mixed state is defined as
\begin{equation}
    \Phi_{k}^{dyn}(t) = -\int_{0}^{t}\mathrm{Tr}[\rho(s)H_{k}(t_{1}^{f})]ds
\end{equation}
Selecting the eigenstates of the post-quench Hamiltonian as a set of basis, i.e. $\{|u_{k\nu}^{f}\rangle\}, \nu = 1, \cdots, m$, we obtain
\begin{equation}
    \begin{split}
        \Phi_{k}^{dyn}(t) & = -\int_{0}^{t}\sum_{\nu=1}^{m}\langle u_{k\nu}^{f}|\rho_{k}(s)H_{k}(t_{1}^{f})|u_{k\nu}^{f}\rangle ds \\
        & = -\sum_{\mu=1}^{m}f_{k\mu}\sum_{\nu=1}^{m}|p_{k\nu}|^{2}\varepsilon_{k\nu}^{f}t.
    \end{split}
\end{equation}

As a result, the Pancharatnam geometric phase for both the pure and mixed states following the quench can be obtained by
\begin{equation}
    \phi_{k}^{G}(t) = \mathrm{Arg}[\mathcal{G}_{k}(t)] - \phi_{k}^{dyn}(t).
\end{equation}

\section{DQPTs in SSH-3 model}
\label{sec:pure.ssh3}

\begin{figure}
    \centering
    \includegraphics[width=1\linewidth]{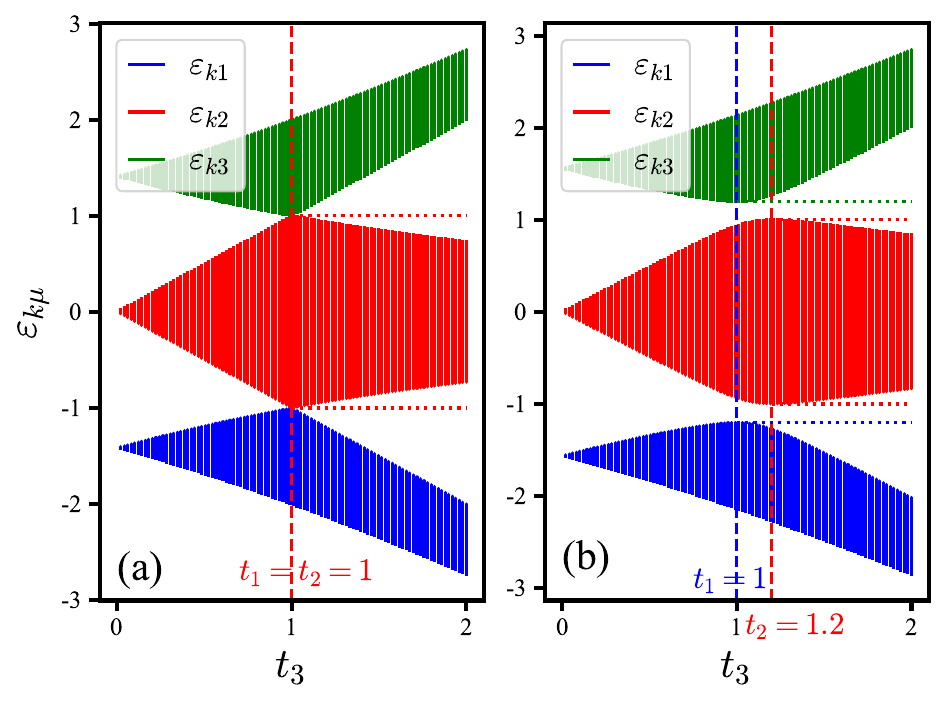}
    \caption{ Energy spectra as a function of the intercell coupling $t_{3}$ in the SSH-3 model for (a) with $t_{1} = t_{2} = 1$ (inversion-symmetric case), and for (b) with $t_{1} = 1, t_{2} = 1.2$ (inversion-symmetric broken case). The solid lines denote the bulk states, and the dotted lines represent the edge states. In (a), the energy gaps close at point $t_{3} = t_{1} = t_{2} = 1$. }
    \label{fig:spectra.ssh3}
\end{figure}

\begin{figure*}
    \centering
    \includegraphics[width=1\linewidth]{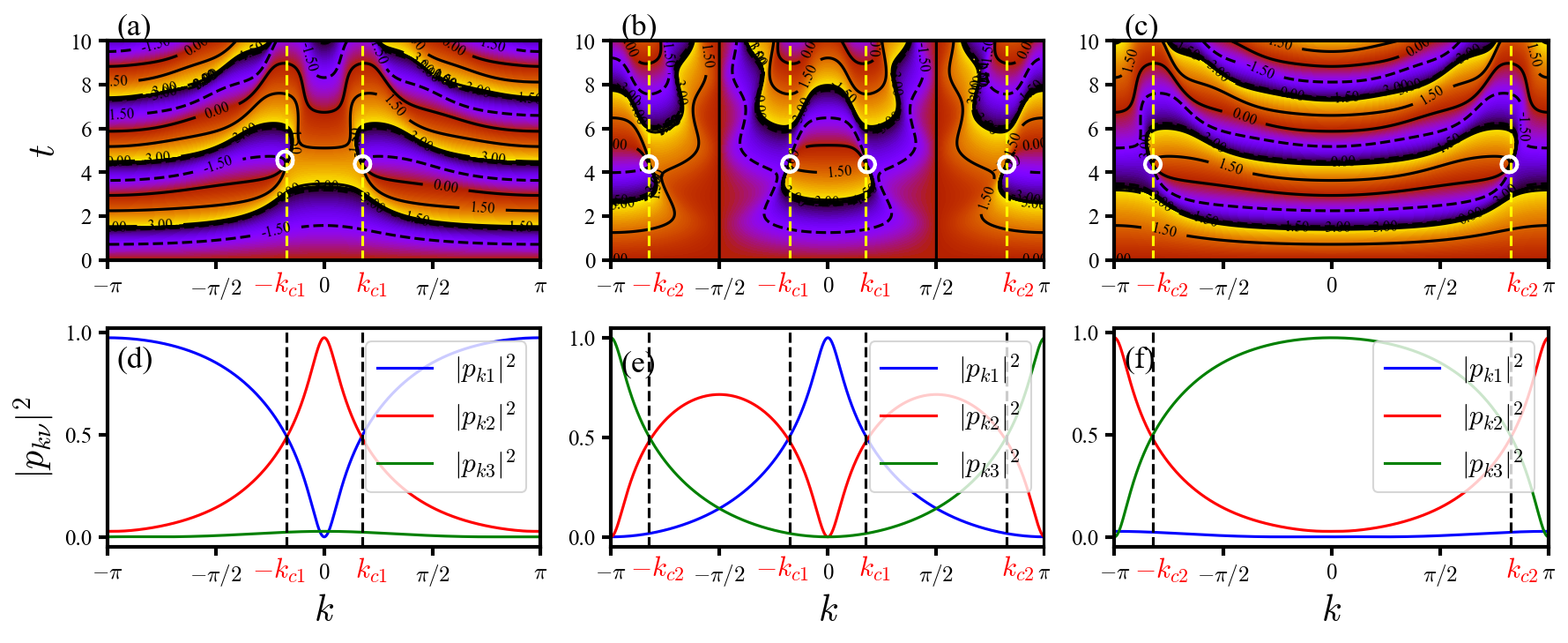}
    \caption{ (a)-(c) The Pancharatnam geometric phases in the SSH-3 model with inversion symmetry from different initial states ($|u_{k1}^{i}\rangle$, $|u_{k2}^{i}\rangle$, and $|u_{k3}^{i}\rangle$ ) are shown in momentum-time space, respectively. The quench paths are from $t_{3}^{i}=0.2$ to $t_{3}^{f}=1.1$ with fixed $t_{1}=t_{2}=1$. The dynamical vortices marked by in white circles exhibit the critical wave vectors and times. (d)-(f) The corresponding transition probabilities $|p_{k\nu}|^{2}$ for different initial states. }
    \label{fig:trans.pgp.ssh3}
\end{figure*}

In this section, we study the DQPTs for pure initial states in the general SSH$-m$ model with $m$ odd, focusing specifically on the SSH-3 model as a representative example. In contrast to the original SSH model, the spectrum of the bulk Hamiltonian for the SSH-3 model is not symmetric about zero energy, indicating the absence of true chiral symmetry \cite{Anastasiadis2022prb, Martinez2019pra}. Chiral symmetry, also known as sublattice symmetry, requires the bulk Hamiltonian to satisfy $\Gamma_{k}H_{k}\Gamma_{k}^{\dag} = -H_{k}$, where $\Gamma$ is a unitary and Hermitian operator. This symmetry ensures the energy spectrum is symmetric around zero, meaning that for any state with energy $E$, there is a chiral symmetric partner with energy $-E$. However, the SSH-3 has an additional symmetry, termed point chiral symmetry, where for each energy $E(k)$ there exists a corresponding state at momentum $\pi-k$ with opposite energy, expressed as $\Gamma_{k}H_{k}\Gamma_{k}^{\dag} = -H_{\pi+k}$. Additionally, the SSH-3 model exhibits inversion symmetry when two of the three couplings are equal. This inversion symmetry is broken when all couplings differ. Nevertheless, edge states persist in both the inversion-symmetric and inversion-symmetry-broken regimes [see Fig.~\ref{fig:spectra.ssh3}].

In Fig.~\ref{fig:spectra.ssh3}, we depict the energy spectra as a function of the intercell coupling $t_{3}$ for systems with and without inversion symmetry. In the inversion-symmetric case, the system exhibits a band-gap closing only when $t_{1} = t_{2} = t_{3}$. In the inversion-symmetric broken case, the system presents a pair of edge states for $t_{1} < t_{3} < t_{2}$, and two pairs of edge states for $t_{3} > t_{2}$. Notably, since the energy gaps remain open throughout the inversion-symmetry-broken regime, the shift in the number of edge states is not driven by the quantum phase transition. We will next explore how the properties of DQPTs are influenced by these different symmetry scenarios.

\subsection{DQPTs for the inversion-symmetric case}
\label{sec.pure.ssh3.gapless}

\begin{figure}
    \centering
    \includegraphics[width=1\linewidth]{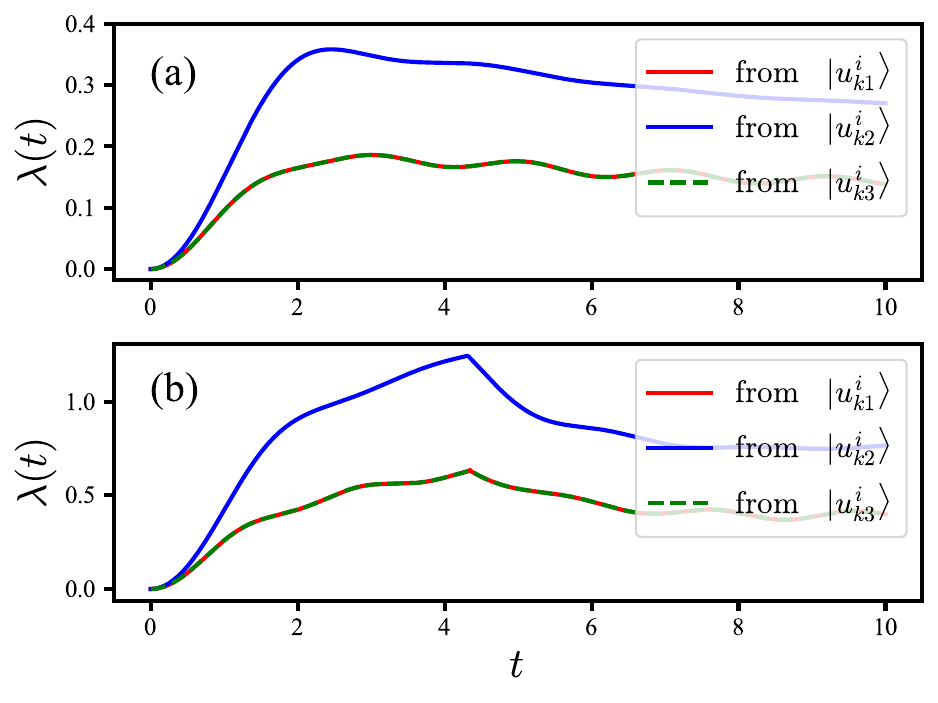}
    \caption{ The rate functions in the SSH-3 models with inversion symmetry for the quenches from different initial states, where the quenches in (a) are from $t_{3}^{i} = 0.2$ to $t_{3}^{f} = 0.8$, and in (b) are from $t_{3}^{i} = 0.2$ to $t_{3}^{f} = 1.1$.}
    \label{fig:rate.ssh3}
\end{figure}

Firstly, we investigate the properties of DQPTs in the SSH-3 model with inversion symmetry, of which the energy spectra show a band-gap closing at the critical point $t_{3c} = 1$ [see Fig.~\ref{fig:spectra.ssh3}~(a)]. In this context, the quench across the critical point of the quantum phase transition is generally regarded as a key condition for the occurrence of DQPT, although some counterexamples have been reported in certain systems \cite{Vajna2014prb, Cao2022cpb}. It is also worth noting that while the Fisher zero theory has traditionally served as a powerful and straightforward tool to identify DQPTs, its application becomes challenging in our model with multiple bands ($m \ge 3$) due to the complexity of solving the equation $\mathcal{G}_{k}(z) = 0, z = \tau + it$. To circumvent this difficulty, we instead employ the concept of the Pancharatnam geometric phase, which reveals dynamical vortices at the critical momenta and critical times [see Figs.\ref{fig:trans.pgp.ssh3}~(a)-(c)].

Since the bulk Hamiltonian of the SSH-3 model possesses three pure eigenstates, $|u_{k\mu}\rangle$ with $\mu = 1, 2, 3$, we analyze quenches starting from different initial pure states. Figures~\ref{fig:rate.ssh3} display typical results of the rate functions corresponding to these quenches. In Figs.~\ref{fig:rate.ssh3}~(a), where the quenches do not cross the critical point $t_{3c}$, the rate functions remain smooth. In contrast, Fig.~\ref{fig:rate.ssh3}~(b) shows that the rate functions for quenches crossing $t_{3c}$ exhibit the cusp-like singularities at the critical time. This indicates that, in the inversion-symmetric system, the emergence of DQPT requires the quench to cross the critical point. Meanwhile, we observe that the rate functions for quenches starting from  $|u_{k1}^{i}\rangle$ and $|u_{k3}^{i}\rangle$ coincide, which can be attributed to the underlying point chiral symmetry of the model.

Furthermore, we can obtain more information on the influence of the point chiral symmetry by analyzing details of the Pancharatnam geometric phases and the transition probabilities. Figs.~\ref{fig:trans.pgp.ssh3}~(a)-(c) depict the Pancharatnam geometric phases in the ($k, t$) plane for quenches from different initial states, and their corresponding transition probabilities $|p_{k\nu}|^{2}$ with $p_{k\nu} = \langle u_{k\nu}^{f}|u_{k\mu}^{i}\rangle$ are presented in (d)-(e). First of all, we find that the cases of quenches from $|u_{k1}^{i}\rangle$ and $|u_{k3}^{i}\rangle$ have one pair of critical wave vectors ($\pm k_{c1}$ and $\pm k_{c2}$), and the case of quench from $|u_{k2}^{i}\rangle$ has two pairs of critical wave vectors. These critical wave vectors correspond to the same critical time and satisfy the following relation:
\begin{equation}
  k_{c1} + k_{c2} = \pi,  
\end{equation}
which matches the property of point chiral symmetry perfectly. Next, we notice that the critical wave vectors are exactly the intersection of the transition probabilities. For instance, in the case of quench from $|u_{k1}^{i}\rangle$, the lines $|p_{k1}|^{2}$ and  $|p_{k2}|^{2}$ intersect at $\pm k_{c1}$; in the case of quench from $|u_{k3}^{i}\rangle$,  the lines $|p_{k2}|^{2}$ and  $|p_{k3}|^{2}$ intersect at $\pm k_{c2}$. This reminds us that, in two-band models, the critical wave vector $k_{c}$ is also closely related to the equal transition probabilities \cite{Cao2024jpcm}, i.e. $|p_{k1}|^{2} = |p_{k2}|^{2} = 1/2$.
The criterion can now be extended to multiband systems: When a DQPT occurs, the critical wave vector can be obtained by the intersection between two transition probabilities that correspond to the two neighbor energy bands.

\subsection{DQPTs for the inversion-symmetric broken case}
\label{sec:pure.ssh3.gap}

\begin{figure}
    \centering
    \includegraphics[width=1\linewidth]{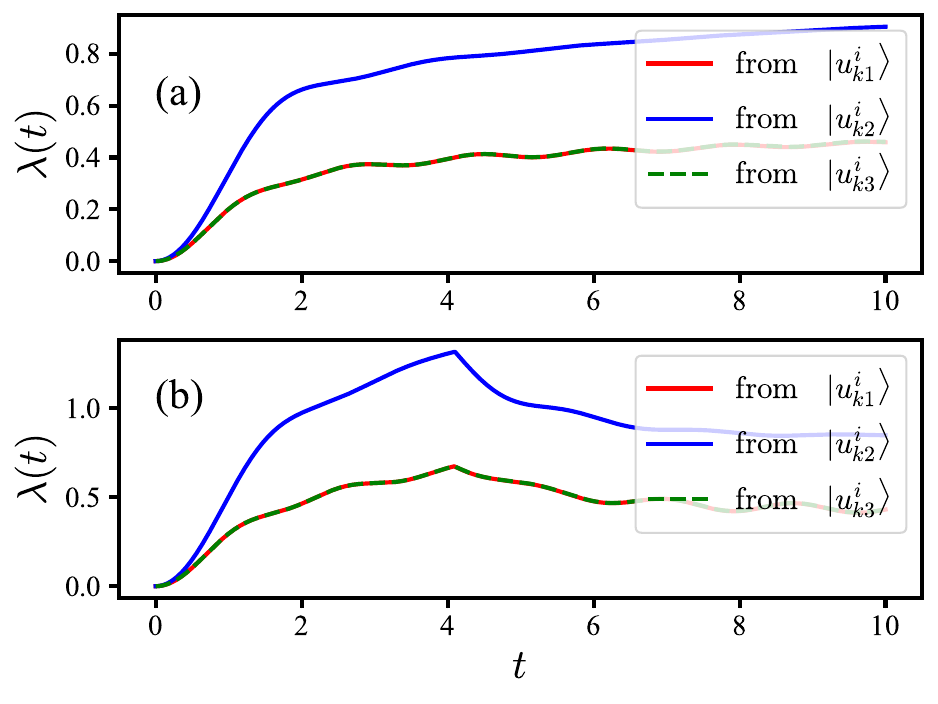}
    \caption{ The rate functions in the SSH-3 models with broken inversion symmetry for the quenches from different initial states, where the quench in (a) is from $t_{3}^{i} = 0.2$ to $t_{3}^{f} = 1.1$ and in (b) is from $t_{3}^{i} = 0.2$ to $t_{3}^{f} = 1.3$.}
    \label{fig:rate.ssh3.gapped}
\end{figure}

\begin{figure}
    \centering
    \includegraphics[width=1\linewidth]{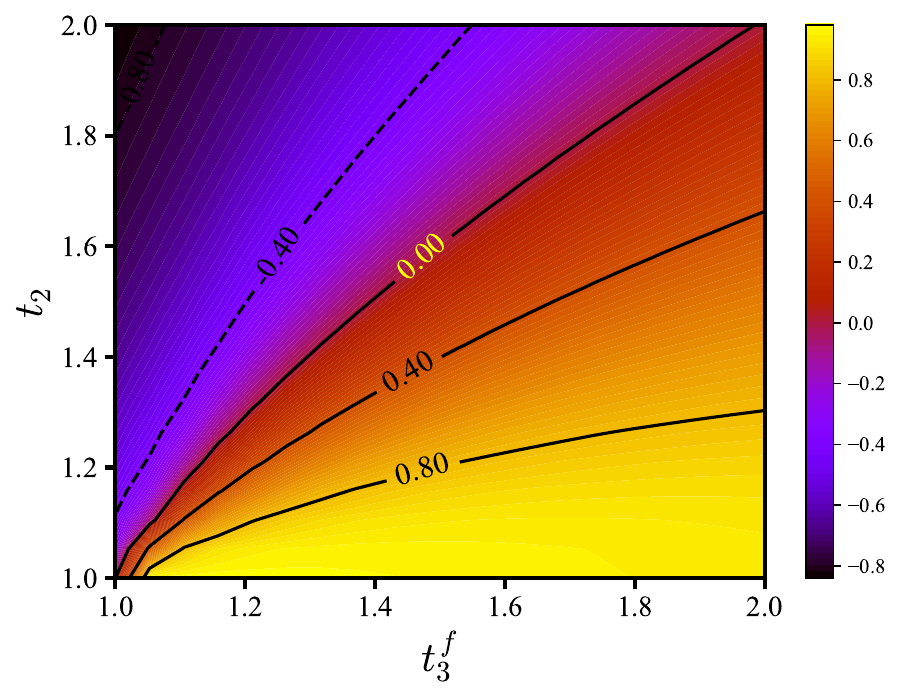}
    \caption{The value of $\mathrm{max}(|p_{k2}|^{2}) - \mathrm{min}(|p_{k1}|^{2})$ in the SSH-3 model with broken inversion symmetry for quenches from the initial state $|u_{k1}^{i}\rangle$. Here, we set the intracell coupling $t_{1} = 1$ and the initial intercell coupling $t_{3}^{i} = 0.2$. The critical line of the couplings $t_{2}$ and $t_{3}^{f}$ is denoted by the line $\mathrm{max}(|p_{k2}|^{2}) - \mathrm{min}(|p_{k1}|^{2}) = 0$. }
    \label{fig:diff.ssh3}
\end{figure}

We now turn to the study of DQPTs in the SSH-3 model without inversion symmetry, where the energy bands remain gapped throughout [see Fig.~\ref{fig:spectra.ssh3}~(b)]. Previous studies have generally held that the presence of an energy gap prevents the occurrence of DQPTs in such systems \cite{Huang2016prl, Cao2024jpcm}. However, the SSH-3 model presents a distinct scenario. As the intercell coupling $t_{3}$ increases, the system experiences two discrete changes in the number of edge states. Notably, these changes do not correspond to phase transitions, since the energy gap remains finite at all times. This unique feature significantly influences the system's nonequilibrium dynamical behavior.

In Fig.~\ref{fig:rate.ssh3.gapped}, we depict the rate functions for quenches starting from various initial states, with intracell couplings fixed at $t_{1} = 1$ and $t_{2} = 1.2$. In Fig.~\ref{fig:rate.ssh3.gapped}~(a), the quench path is from $t_{3}^{i} = 0.2$ to $t_{3}^{f} = 1.1$, crossing the first point at $t_{3} = 1$ where the number of edge states changes. The resulting smooth rate functions indicate the absence of DQPTs in this regime.  In contrast, Fig.~\ref{fig:rate.ssh3.gapped}~(b) shows a quench to $t_{3}^{f} = 1.3$, which crosses the second such critical point at $t_{3} = 1.2$. Here, the rate function displays a pronounced cusp-like singularity, signaling the occurrence of a DQPT. These findings suggest the existence of a critical value $t_{3c}$  such that when the quench crosses this threshold ($t_{3}^{f}>t_{3c}$), a DQPT can emerge even though the system remains gapped.

To determine the critical threshold $t_{3c}$ for the presence of DQPT, we can utilize the condition derived in the last section \ref{sec.pure.ssh3.gapless}: the critical wave vectors are exactly the intersection of the transition probabilities. For a quench starting from the initial pure state $|u_{k1}^{i}\rangle$, this condition can be expressed as
\begin{equation}
    \mathrm{min}(|p_{k1}|^{2}) < \mathrm{max}(|p_{k2}|^{2}).
\end{equation}
At the critical threshold $t_{3}^{f}$, this inequality becomes an equality:
\begin{equation}
    \mathrm{min}(|p_{k1}|^{2}) = \mathrm{max}(|p_{k2}|^{2}).
\end{equation}
Furthermore, as illustrated in Fig.~\ref{fig:rate.ssh3.gapped}~(b), the critical threshold $t_{3c}$ is strongly related to the intracell coupling $t_{2}$. 
To explore this relationship, Fig.~\ref{fig:diff.ssh3} shows the variation of $\mathrm{max}(|p_{k2}|^{2}) - \mathrm{min}(|p_{k1}|^{2}) = 0$ as a function of $t_{2}$ and $t_{3}^{f}$ for quenches starting from $|u_{k1}^{i}\rangle$, with fixed parameters $t_{1} = 1$ and $t_{3}^{i} = 0.2$. The critical threshold $t_{3c}$ is identified by the contour line 
\begin{equation}
    \mathrm{max}(|p_{k2}|^{2}) - \mathrm{min}(|p_{k1}|^{2}) = 0.
\end{equation}
From the plot, it is evident that $t_{3c}$ increases with $t_{2}$, though the dependence is nonlinear. More specifically, $t_{3c}$ is generally slightly larger than $t_{2}$, except at the point $t_{2} = 1$. At $t_{2} = 1$, the system reduces to the SSH-3 with inversion symmetry, for which the threshold is exactly $t_{3c} = 1$. This observation aligns well with previous results: in the inversion-symmetric regime, a DQPT occurs only when the quench crosses the system’s critical point.

\subsection{DQPT for mixed states in SSH-3 model}
\label{sec:ssh3.mixed}

\begin{figure}
    \centering
    \includegraphics[width=1\linewidth]{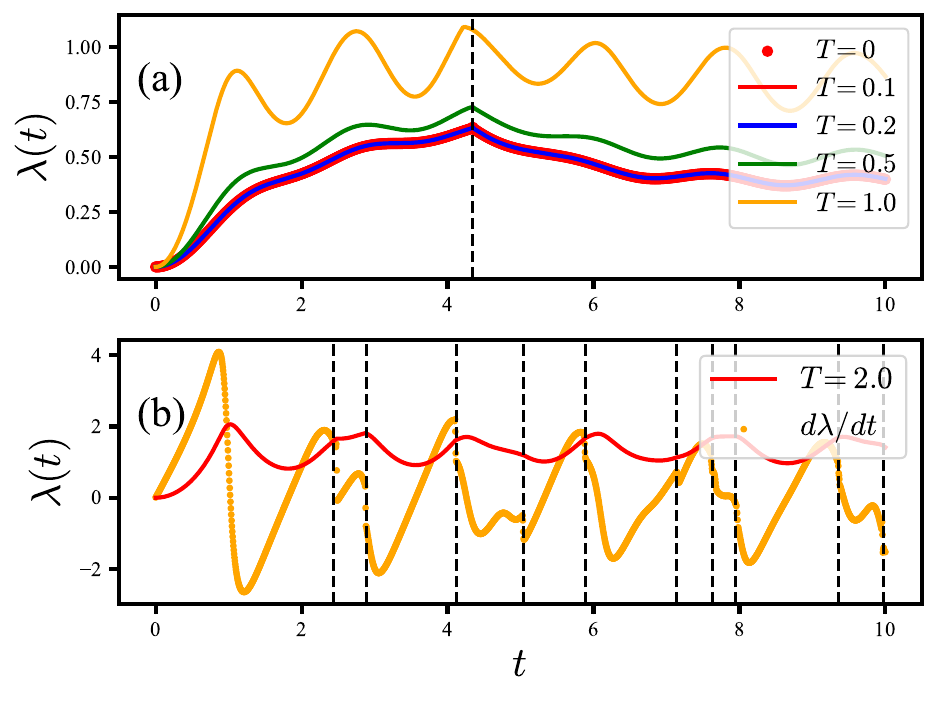}
    \caption{(a) The rate functions in the SSH-3 model for mixed states at different temperatures ($T=1/\beta$), where the quenches are all from $t_{3}^{i} = 0.2$ to $t_{3}^{f} = 1.1$, with fixed $t_{1} = t_{2} = 1$. The red-dot line denotes the rate function for the initial state $|u_{k1}^{i}\rangle$. (b) The rate function and its derivative  $d\lambda/dt$ for the initial state at high $T = 2.0$. The later is used to highlight the critical times more clearly. The dashed lines mark the critical times.}
    \label{fig:rate.ssh3.mixed}
\end{figure}

In this section, we investigate the influence of the finite temperature on the DQPT in the SSH-3 model, in which the dynamics is described by the generalized Loschmidt amplitude for a quench from mixed initial states. First, it is necessary to clarify that since we find similar behavior of mixed-state DQPTs in the SSH-3 model both with and without inversion symmetry, the following results will use the system with inversion symmetry as an example to illustrate the nature of DQPTs at finite temperature.

\begin{figure}
    \centering
    \includegraphics[width=1\linewidth]{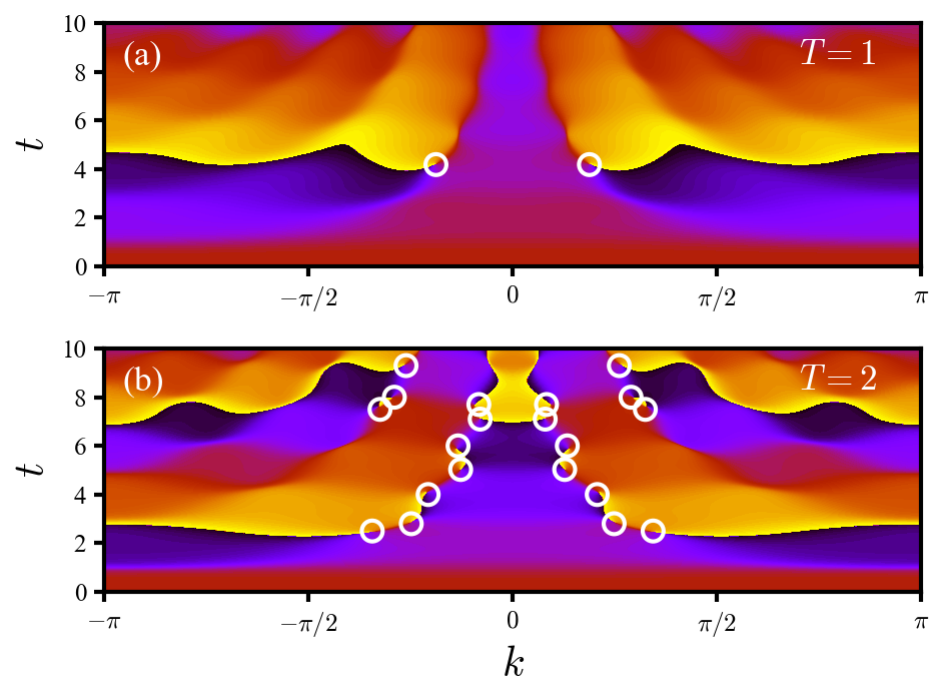}
    \caption{The Pancharatnam geometric phases in the SSH-3 model for initial states at different temperatures, (a) $T=1$, and (b) $T=2$. The quench paths are from $t_{3}^{i} = 0.2$ to $t_{3}^{f} = 1.1$, with fixed $t_{1} = t_{2} = 1$. The dynamical vortices are highlighted in white.}
    \label{fig:pgp.ssh3.mixed}
\end{figure}

Fig.~\ref{fig:rate.ssh3.mixed} shows the rate functions in the SSH-3 model for initial states at different temperatures ($T = 1/\beta$). It can be observed that at very low temperatures, i.e. $T = 0.1$ and $T = 0.2$, the rate function almost coincides with the result of that starting from the initial pure states $|u_{k1}^{i}\rangle$. This indicates that the behavior of the DQPT is not affected at low temperatures. As the temperature increases, the critical time will gradually deviate from the critical time at pure states and lower temperatures. Especially, at $T = 0.5$, the numerical values of the rate function change but the critical time still equals that at lower temperatures. However, when the temperature is very high, the nature of DQPTs will undergo significant changes, to the extent that we observe a large number of critical times emerging in the system through the first derivative $d\lambda/dt$ of the rate function. In addition, the results of DQPTs are also confirmed through the Pancharatnam geometric phase (see Fig.~\ref{fig:pgp.ssh3.mixed}). Especially, at high temperature $T = 2.0$, the Pancharatnam geometric phase exhibits many dynamical vortices emerging in the ($k-t$) plane.

The behavior of the DQPT for mixed states at finite temperatures can be explained based on the properties of that from pure states. At lower temperatures, the probabilities $f_{k\mu}$ of the thermal states satisfy 
\begin{equation}
    f_{k1} \gg f_{k2} > f_{k3},
\end{equation}
thus the generalized Loschmidt amplitude is approximately equal to the case starting from the lowest energy band in the pure state [see results of $T=0.1, 0.2$ in Fig.~\ref{fig:pgp.ssh3.mixed}~(a)]. As the temperature increases, the probabilities $f_{k\mu}$ of the thermal states shift to
\begin{equation}
   f_{k1} > f_{k2} \gg f_{k3},   
\end{equation}
so that the generalized Loschmidt amplitude is approximately equal to the cases only mixing the quench from pure states $|u_{k1}^{i}\rangle$ and $|u_{k2}^{i}\rangle$. According to Fig.~\ref{fig:trans.pgp.ssh3}, we know that the cases of quenches from initial states $|u_{k1}^{i}\rangle$ and $|u_{k2}^{i}\rangle$ have the common critical wave vectors $k_{c1}$ and $-k_{c1}$. In this case, the critical times for mixed states still do not change or just deviate slightly from that at lower temperatures [see results of $T=0.5, 1.0$ in Fig.~\ref{fig:pgp.ssh3.mixed}~(a)]. However, at high temperatures, due to the contribution of that from the highest band $|u_{k3}^{i}\rangle$ no longer being negligible. Considering the cases of quenches from the initial pure states $|u_{k2}^{i}\rangle$ and $|u_{k3}^{i}\rangle$ have another common critical time $k_{c2} (-k_{c2})$, distinct from $k_{c1}$, the behavior of the DQPT at high temperatures is significantly changed [see results of $T=2.0$ in Fig.~\ref{fig:pgp.ssh3.mixed}~(b)].

\section{DQPTs in SSH-4 model}
\label{sec:pure.ssh4}

\begin{figure}
    \centering
    \includegraphics[width=1\linewidth]{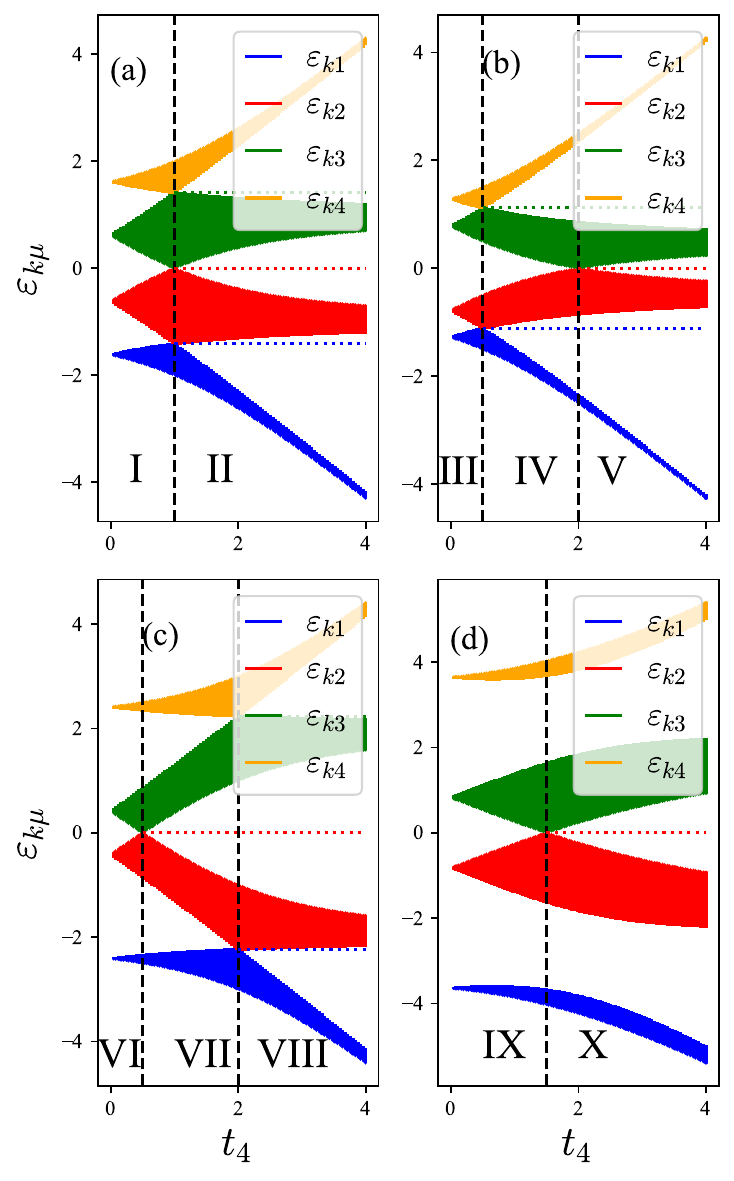}
    \caption{Energy spectra as a function of the intercell coupling $t_{4}$ in the SSH-4 models, for (a) with the intracell couplings $t_{1} = t_{2} = t_{3} = 1$, for (b) with $t_{1} = t_{3} = 1$ and $t_{2} = 0.5$, for (c) with $t_{1} =  t_{3} = 2$ and $t_{2} = 1$, and for (d) with $t_{1} = 1, t_{2} = 2, t_{3} = 3$. The solid lines denotes the bulk states, the dotted lines represent the edge states. The nonzero edge modes are given by $\varepsilon_{k} = \pm\sqrt{t_{1}^{2}+t_{2}^{2}}$. For convenience, we label the phases using Roman letters.}
    \label{fig:energy.ssh4}
\end{figure}

In this section, we study the DQPTs in the generalized SSH$-m$ model with $m$ even, taking the SSH-4 model as a typical example. In contrast to the SSH$-m$ model with $m$ odd, the bulk Hamiltonian of the SSH-4 still possesses the chiral symmetry \cite{Lee2022cjp}, i.e. $\Gamma_{k}H_{k}\Gamma_{k}^{\dag} = -H_{k}$ with $\Gamma = \mathrm{diag}(1, -1, 1, -1)$. 

By calculating the energy spectra under different configurations of couplings ($t_{1}, t_{2}, t_{3}, t_{4}$), we identify three typical energy spectrum configurations, characterized by the different locations of edge states, in the SSH-4 model. If all the intracell couplings are equal, e.g. $t_{1} = t_{2} = t_{3} = 1$, the Hamiltonian exhibits the band-gap closing between two neighboring energy spectra at the critical point $t_{4c} = 1$ [see Fig.~\ref{fig:energy.ssh4}~(a)]. Three edge modes (one zero edge mode and two nonzero edge modes $\varepsilon_{k} = \pm\sqrt{t_{1}^{2}+t_{2}^{2}}$) are present in the phase with $t_{4} > 1$. If the intracell couplings satisfy $t_{1} = t_{3} \neq t_{2}$, the Hamiltonian has two critical points $t_{4c}$, at which the band gap closes [see Figs.~\ref{fig:energy.ssh4}~(b) and (c)]. The two critical points are obtained by
\begin{equation}
    t_{4c} = \frac{t_{1}t_{3}}{t_{2}}, \quad \frac{t_{2}t_{3}}{t_{1}},
\end{equation}
respectively. In this case, the locations of edge states are determined by $t_{2}$: if $t_{2} < t_{1} (t_{3})$, the Hamiltonian has two nonzero edge modes for $\frac{t_{2}t_{3}}{t_{1}}<t_{4}<\frac{t_{1}t_{3}}{t_{2}}$, and three edge modes (one zero edge mode and two nonzero edge modes) for $t_{4}>\frac{t_{1}t_{3}}{t_{2}}$ [see Fig.~\ref{fig:energy.ssh4}~(b)]; if $t_{2} > t_{1} (t_{3})$, the Hamiltonian has one zero edge mode for $\frac{t_{1}t_{3}}{t_{2}}<t_{4}<\frac{t_{2}t_{3}}{t_{1}}$, and three edge modes (one zero edge mode and two nonzero edge modes) for $t_{4}>\frac{t_{2}t_{3}}{t_{1}}$ [see Fig.~\ref{fig:energy.ssh4}~(c)]. Finally, if the intracell couplings satisfy $t_{1} \neq t_{3}$, the Hamiltonian has only one critical point at
\begin{equation}
    t_{4c} = \frac{t_{1}t_{3}}{t_{2}},
\end{equation}
and one zero edge mode is present for $t_{4} > \frac{t_{1}t_{3}}{t_{2}}$.

\subsection{DQPTs for pure states in SSH-4 model}

\begin{figure}
    \centering
    \includegraphics[width=1\linewidth]{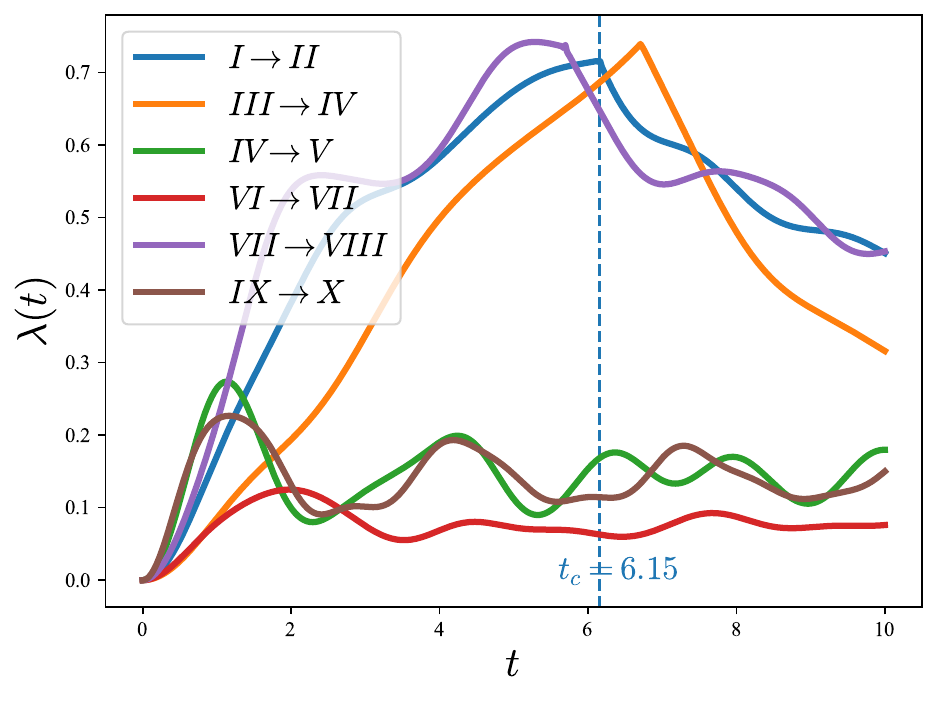}
    \caption{Rate functions in the SSH-4 model with different energy spectra configurations for the quench from the initial pure states $|u_{k1}^{i}\rangle (|u_{k4}^{i}\rangle)$.}
    \label{fig:rate.ssh4.1}
\end{figure}

\begin{figure}
    \centering
    \includegraphics[width=1\linewidth]{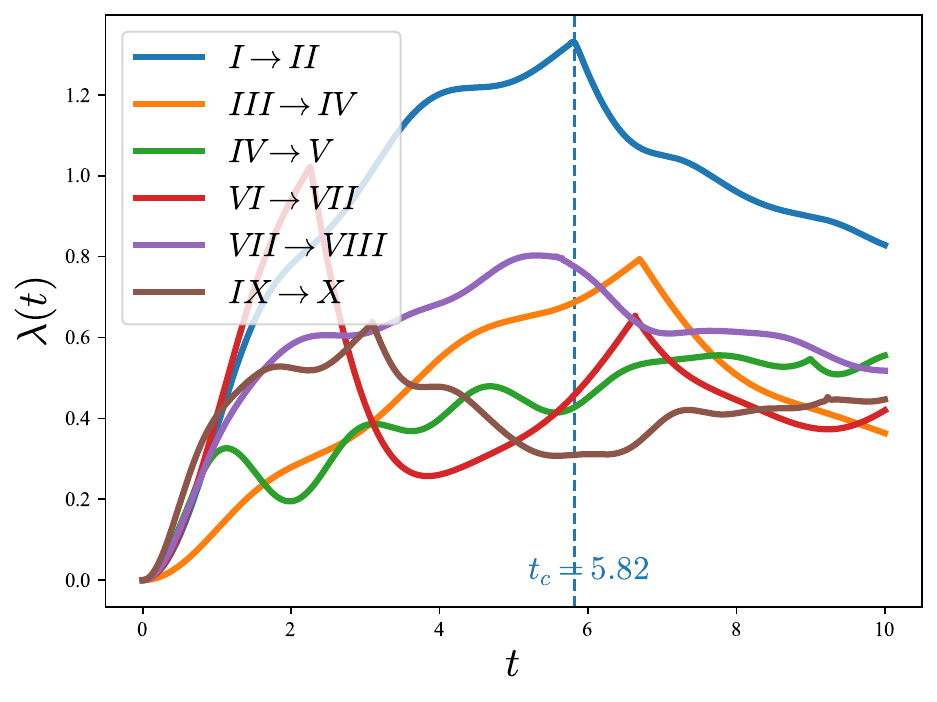}
    \caption{Rate functions in the SSH-4 model with different energy spectra configurations for the quench from the initial pure states $|u_{k2}^{i}\rangle (|u_{k3}^{i}\rangle)$.}
    \label{fig:rate.ssh4.2}
\end{figure}

Now we investigate the DQPTs in the SSH-4 model from the pure states. First, it should be noted that, similar to the Kitaev chain which possesses particle-hole symmetry \cite{Cao2024jpcm}, the SSH-4 model exhibits chiral symmetry, resulting in the system's energy spectrum satisfying $ \varepsilon_{k1} = -\varepsilon_{k4}$ and $\varepsilon_{k2} = -\varepsilon_{k3}$. Moreover, along the same quench pathway, the rate functions obtained from quantum quenches starting from $|u_{k1}^{i}\rangle$ and $|u_{k4}^{i}\rangle$, as well as from $|u_{k2}^{i}\rangle$ and $|u_{k3}^{i}\rangle$, are exactly the same. Therefore, here we only need to consider the results starting from $|u_{k1}^{i}\rangle$ and $|u_{k2}^{i}\rangle$.

Figs.~\ref{fig:rate.ssh4.1} show the typical results of rate functions for various types of quenches starting from the initial pure states $|u_{k1}^{i}\rangle$ in the SSH-4 model, under different energy spectrum configurations. We observe that quenches from phase I to phase II, phase III to phase IV, and phase VII to phase VIII exhibit cusp-like singularities in the rate functions, indicating the presence of dynamical quantum phase transitions (DQPTs). In contrast, other quenches crossing critical points show smooth rate functions without DQPT signatures. Notably, the quenches accompanied by DQPTs correspond to Hamiltonians where the energy gap closes between bands $\varepsilon_{k1}$ and $\varepsilon_{k2}$. These results suggest two necessary conditions for the occurrence of DQPTs: (1) the energy band of the initial state must have a gapless point with other bands, and (2) the quench must cross this gapless point. To verify this, Fig.~\ref{fig:rate.ssh4.2} shows the rate functions for the same quenches starting from the initial states $|u_{k2}^{i}\rangle$ in Fig.~\ref{fig:rate.ssh4.2}. As illustrated in Fig.~\ref{fig:energy.ssh4}, critical points always exist where the energy gap between $\varepsilon_{k2}$ and other bands ($\varepsilon_{k1}$ or $\varepsilon_{k3}$) closes. Consequently, we observe that the quenches from $|u_{k2}^{i}\rangle$ exhibit cusp-like singularities, confirming the essential role of gap closing in DQPT emergence.

\begin{figure}
    \centering
    \includegraphics[width=1\linewidth]{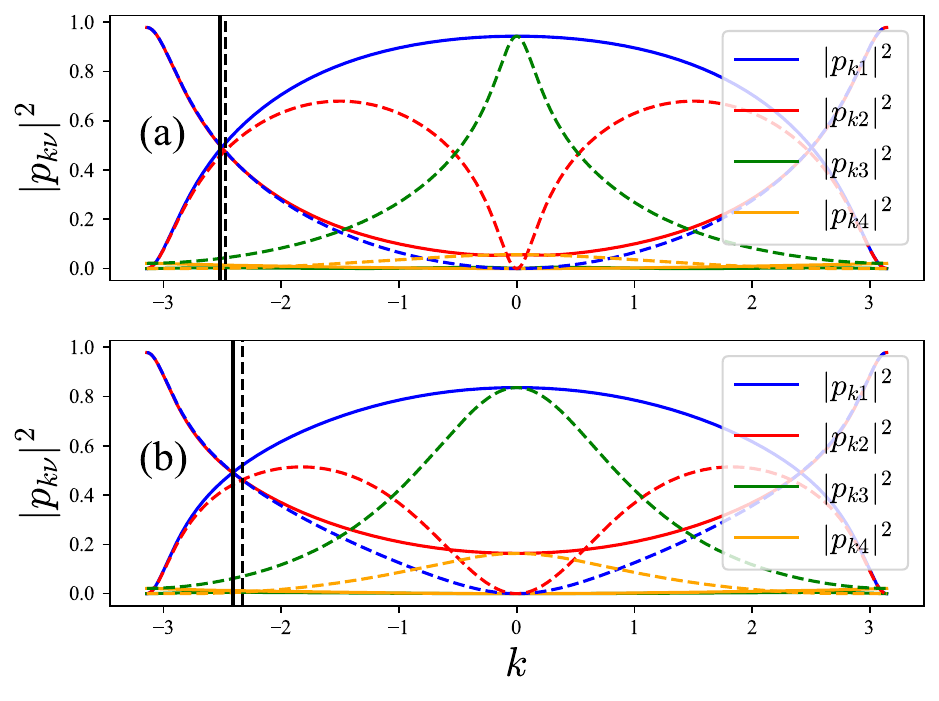}
    \caption{The transition probabilities $|p_{k\nu}|^{2}$ in the SSH-4 model. The results in (a) correspond to the quench path of Fig.~\ref{fig:rate.ssh4.1}~(b), and in (b) correspond to that of Fig.~\ref{fig:rate.ssh4.2}~(b). The solid lines denote the results of the quench from the initial state $|u_{k1}^{i}\rangle$, and the dashed lines denote the results of the quench from the initial state $|u_{k2}^{i}\rangle$. }
    \label{fig:transition.probability.ssh4}
\end{figure}

Some details in the results are also worth noting. In the SSH-3 model, we observe that the critical times obtained from quenches starting from different initial states are the same. However, in the SSH-4 model, the situation is different. From Figs.~\ref{fig:rate.ssh4.1} and \ref{fig:rate.ssh4.2}, it can be seen that the critical times of the rate functions for the quenches starting from $|u_{k1}^{i}\rangle$ and $|u_{k2}^{i}\rangle$ are not equal ($t_{c}=6.15$ for $|u_{k1}^{i}\rangle$ and $t_{c}=5.82$ for $|u_{k2}^{i}\rangle$). This difference is also reflected in the deviation of the critical wave vectors [see Fig.~\ref{fig:spectra.ssh3}~(a) and (b)]. While these details may seem less important in the context of pure-state DQPTs, they will significantly alter the behavior of mixed-state DQPTs.

\subsection{DQPTs for mixed states in SSH-4 model}

\begin{figure}
    \centering
    \includegraphics[width=1\linewidth]{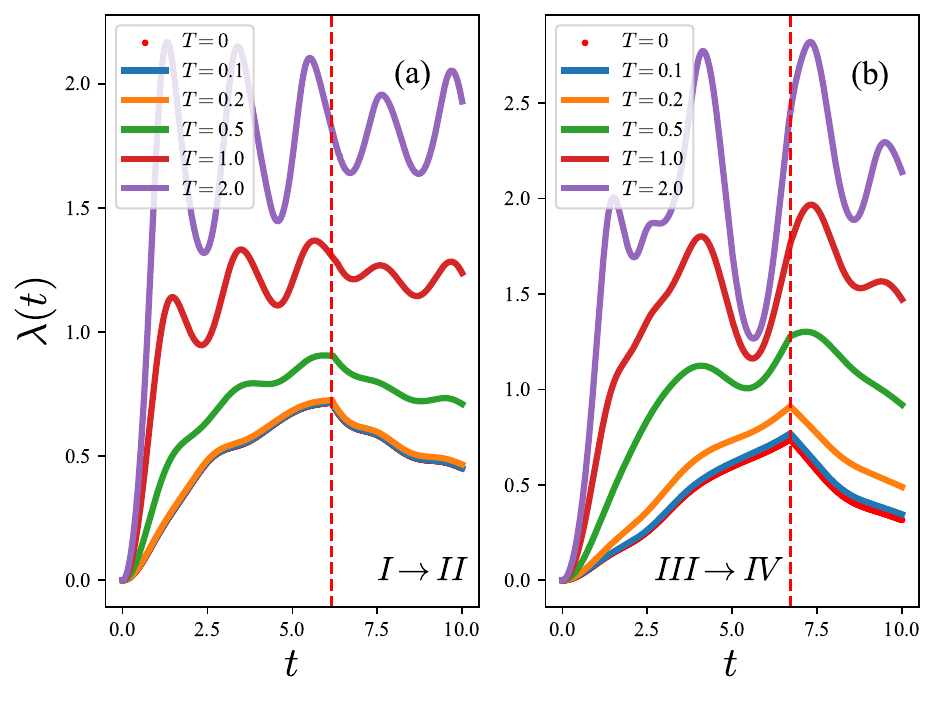}
    \caption{ Rate functions in the SSH-4 model for mixed states at different temperatures. The red-dot line denotes the rate function for the initial state $|u_{k1}^{i}\rangle$ at zero temperature.}
    \label{fig:rate.ssh4.mixed}
\end{figure}

\begin{figure}
    \centering
    \includegraphics[width=1\linewidth]{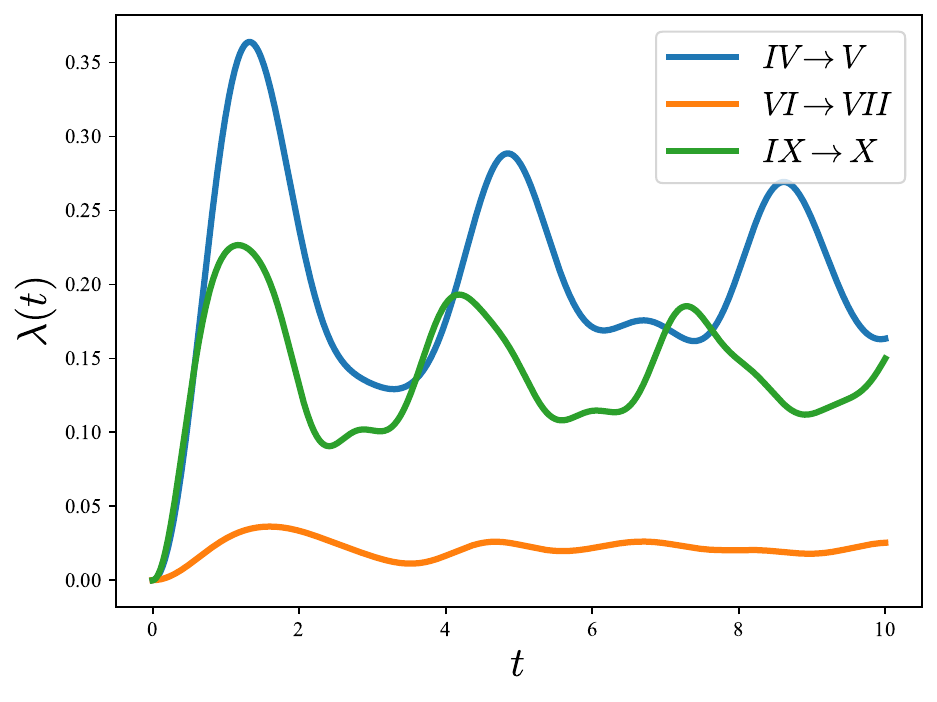}
    \caption{ Rate functions in the SSH-4 models for various types of quenches at $T=0.1$. }
    \label{fig:rate.ssh4.mixed.2}
\end{figure}

We now turn to investigating the influence of finite temperature on DQPTs, where the quench begins from mixed initial states. Broadly, two scenarios arise for DQPTs in mixed states: one in which a quench starting from the pure state $|u_{k1}^{i}\rangle$ induces a DQPT, and another where it does not.

Figs.~\ref{fig:rate.ssh4.mixed} (a) and (b) show the rate functions at various temperatures for quenches from phase I to phase II and from phase III to phase IV, respectively. For comparison, the rate function results for the corresponding initial pure state $|u_{k1}^{i}\rangle$ are also depicted. At relatively low temperatures, the rate functions closely follow those of the pure states. However, as temperature increases, the rate functions become smooth and no longer display non-analytic critical points, indicating the suppression of DQPTs in mixed states at elevated temperatures. 

Similar to observations in the SSH-3 model, the finite-temperature behavior of DQPTs in SSH-4 model for mixed states can be understood by considering the pure-state DQPTs. At low temperatures, the thermodynamic occupation probabilities satisfy
\begin{equation}
    f_{k1} \gg f_{k2}>f_{k3}>f_{k4},
\end{equation}
thus the generalized Loschmidt amplitude of the system is approximately equal to the Loschmidt amplitude from quenching starting from the lowest energy band ($|u_{k1}^{i}\rangle$) closely approximating the pure-state Loschmidt amplitude. Consequently, DQPT behavior at low temperatures mirrors that of the pure state. At higher temperatures, however, the increased population of higher energy bands—each characterized by distinct critical momenta in the quench dynamics [see Fig.~\ref{fig:transition.probability.ssh4} (a) and (b)]—results in a generalized Loschmidt amplitude free of Fisher zeros, thereby eliminating DQPTs in the mixed state.

Conversely, if a quench starting from the pure state $|u_{k1}^{i}\rangle$ does not induce a DQPT, then DQPTs remain absent in the corresponding mixed-state quenches at all temperatures, including low temperature regimes [see Fig.~\ref{fig:rate.ssh4.mixed.2}].

\section{conclusion}
\label{sec:conclusion}

In this paper, we investigate DQPTs for both pure and mixed states within generalized SSH models, focusing on two representative cases: the SSH-3 and SSH-4 models, which differ in their symmetry properties. The SSH-3 model lacks chiral symmetry but possesses a chiral-like, point symmetry that protects robust, localized edge states associated with the system’s topology. Our results show that, for pure states, DQPTs occur following quenches that cross the topological phase transition—even when the energy band gap remains open. For mixed states, DQPT behavior at low temperatures closely follows that of pure states, while at higher temperatures significant modifications arise, including the appearance of multiple critical times. In contrast, the SSH-4 model, which exhibits chiral symmetry, allows exploration of four distinct energy spectrum configurations. We find that pure-state DQPTs require the quench to initiate from an initial state with a gapless band and to cross the critical point of the topological transition. At finite temperature, when a quench from the lowest energy band induces a DQPT, mixed-state DQPTs persist at low temperature but vanish at elevated temperatures. Conversely, if a quench starting from the pure state $|u_{k1}^{i}\rangle$ fails to induce a DQPT, no DQPTs emerge for mixed states at any temperature.

\begin{acknowledgments}
  K. Cao thanks Mrs.~H.~Zhou and Prof.~P.~Tong for the valuable suggestions. The work is supported by the National Natural Science Foundation of China (Grant No. 11875047).
\end{acknowledgments}



\nocite{*}
\bibliography{reference}

\end{document}